# Prototype Active Silicon Sensor in 150 nm HR-CMOS Technology for ATLAS Inner Detector Upgrade

**Piotr Rymaszewski[c], Marlon Barbero[a], Patrick Breugnon[a], Stépahnie Godiot[a], Laura Gonella[c], Tomasz Hemperek[c], Toko Hirono[c], Fabian Hügging[c], Hans Krüger[c], Jian Liu[a], Patrick Pangaud[a], Ivan Peric[b], Alexandre Rozanov[a], Anqing Wang[a], Norbert Wermes[c]**

[a] *Centre de Physique des Particules de Marseille,*
  *Avenue de Luminy 163, Marseille, France*
[b] *Karlsruher Institut für Technologie,*
  *Hermann-von-Helmholtz-Platz 1, Karlsruhe, Germany*
[c] *Physikalisches Institut der Universität Bonn,*
  *Nussallee 12, Bonn, Germany*
  *E-mail:* rymaszewski@physik.uni-bonn.de

ABSTRACT: The LHC Phase-II upgrade will lead to a significant increase in luminosity, which in turn will bring new challenges for the operation of inner tracking detectors. A possible solution is to use active silicon sensors, taking advantage of commercial CMOS technologies. Currently ATLAS R&D programme is qualifying a few commercial technologies in terms of suitability for this task. In this paper a prototype designed in one of them (LFoundry 150 nm process) will be discussed. The chip architecture will be described, including different pixel types incorporated into the design, followed by simulation and measurement results.

# Contents

## 1. Introduction

The LHC Phase-II upgrade will lead to a significant increase in luminosity, which in turn will bring new challenges for the operation of the inner tracking detectors. In order to prepare for this, an ATLAS R&D collaboration called "CMOS Demonstrator" was started with the aim to examine different technologies and choose the most adequate one. One of the possible choices is an active silicon sensor.

Monolithic active pixel sensors (MAPS) allow integration of sensitive sensor volume and signal processing electronics in a single chip, making them an interesting, cost-efficient alternative to hybrid pixel detectors. However, due to relying on diffusion to collect charge, MAPS devices can struggle to meet the timing requirements of detectors operating in high luminosity environments like HL-LHC as shown in [1][2]. This problem can be potentially resolved using modern CMOS technologies to construct depleted MAPS (DMAPS) [3][4][5], which collect charges through the drift in an electric field present in the depletion zone. The depletion width can be calculated as follows (assuming a simple P-N diode model) [6]:

$$d = 0.3 \cdot \sqrt{U_{bias} \cdot \rho_{sub}} \quad (1)$$

where $U_{bias}$ $[V]$ is the reverse bias voltage and $\rho_{sub} [\Omega \cdot cm]$ is the substrate resistivity. This indicates that the process in which the DMAPS structure will be developed should have a high bias voltage capability, and/or the structure should be manufactured on a high resistivity substrate. The discussed prototype, named CCPD_LF, was developed to evaluate the feasibility of using a DMAPS detector based on a high resistivity substrate in the ATLAS experiment. In this paper the device architecture will be presented in detail in Section 2, followed by a discussion of measurement results in Section 3. Plans for the future prototypes will be explained in Section 4, and a summary will be given in Section 5.

## 2. CCPD_LF prototype

The presented prototype was designed using a 150nm CMOS technology (LFoundry [7]). This process has a high voltage option and multiple nested wells (N- and P-wells, deep N- and P-wells, very deep N-wells). The design was manufactured on a high resistive (approx. 2 kΩ·cm) p-type Czochralski silicon.

### 2.1 Sensor structure

For the discussed prototype two sensor architectures, version A and B, were implemented as shown in Figure 1. In both cases the charges created in the P-type bulk by a particle are collected by an N-type well. However, due to different biasing schemes, the two sensor types could not have been included in one matrix. Instead, two completely separate chips (5 mm × 5 mm) were made for this prototype submission, each with its own guardrings and padring. In both chips the pixel size is the same (33 μm × 125 μm), and both pixel matrices have the same number of pixels in them (2736 each). Identical guardring structures surround the matrices (one N-well ring followed by ten P-well rings, the last of which sets the potential level of the substrate), and biasing schemes for versions A and B are summarised in Table 1.

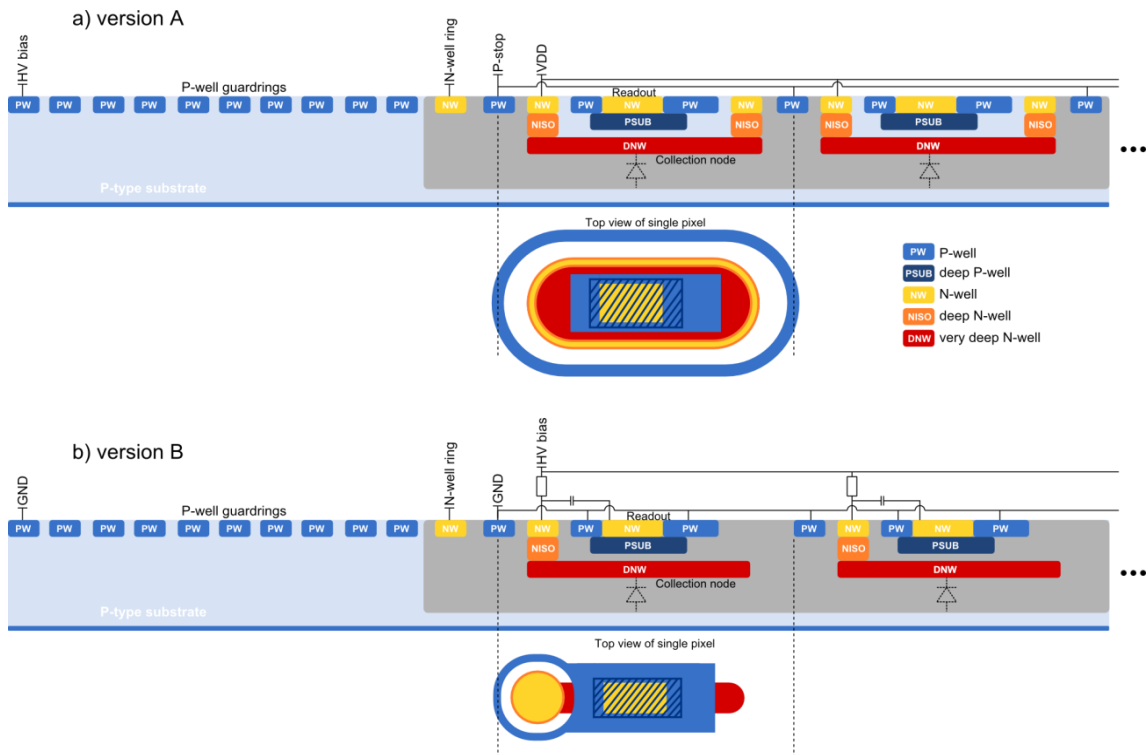

**Figure 1.** Cross sections and top views of CCPD_LF sensors. Depletion region is shown in grey. Diode symbol represents the sensor diode formed by reverse bias voltage.

**Table 1.** Sensor biasing schemes for sensor versions A and B.

|  | **Version A** | **Version B** |
|---|---|---|
| **Readout – substrate isolation** | Isolated by collecting well | Not isolated |
| **Substrate potential** | Negative high voltage | Ground |
| **Potential of collecting well** | 1.8V (power supply for readout) | Positive high voltage |
| **Charge sensitive volume** | P-N junction between collection well and substrate | |

An important difference between the two sensor versions is the size of the charge collecting well, which is approx. double the size in version A. This directly influences the input capacitance to the readout electronics and therefore it is expected that version B will exhibit lower readout noise and that its charge sensitive amplifier (CSA) will have a faster rise time. On the other hand, a smaller collection electrode means a lower fill factor and as a result a lower charge collection efficiency. However, this can potentially be compensated for by, for example, increasing the reverse bias voltage leading to a stronger electric field.

## 2.2 Readout electronics

### 2.2.1 In-pixel circuitry

Figure 2a presents a simplified block diagram of the readout electronics inside a pixel. The sensing element is AC coupled to the input of a charge sensitive amplifier. The CSA (shown in Figure 2b) is a folded cascode amplifier designed to consume 10 µA of bias current, which allows an input equivalent noise of 124e$^-$, a gain of 18 $\mu V/e^-$, and a peaking time of 31.2 ns (for 4000e$^-$ input at an input capacitance of 150fF). The amplified signal is sent through a coupling capacitor (to allow baseline adjustment) to the input of a discriminator. A schematic of the implemented discriminator (two-stage open-loop architecture) is presented in Figure 2c. The discriminator's threshold is set globally for all pixels in the matrix but can be adjusted locally using 4-bit trim DAC to reduce the dispersion. The further signal path depends on the mode of operation of the prototype:

1. Charge coupled pixel detector (CCPD) mode: The prototype matrix is bump bonded or glued to a readout chip (FEI-4). In this case, the digital signal from the discriminator is sent to the output stage (shown in Figure 2c) where its width can be stretched (required for proper operation with FEI-4). After that, the second part of output stage sends a signal of adjustable amplitude through a coupling capacitor to a bond pad.
2. Standalone mode: The hit information from the discriminator is stored in a D flip-flop and later read out via a global shift register. This functionality allows obtaining an image from the sensor without the need to connect an additional readout chip.

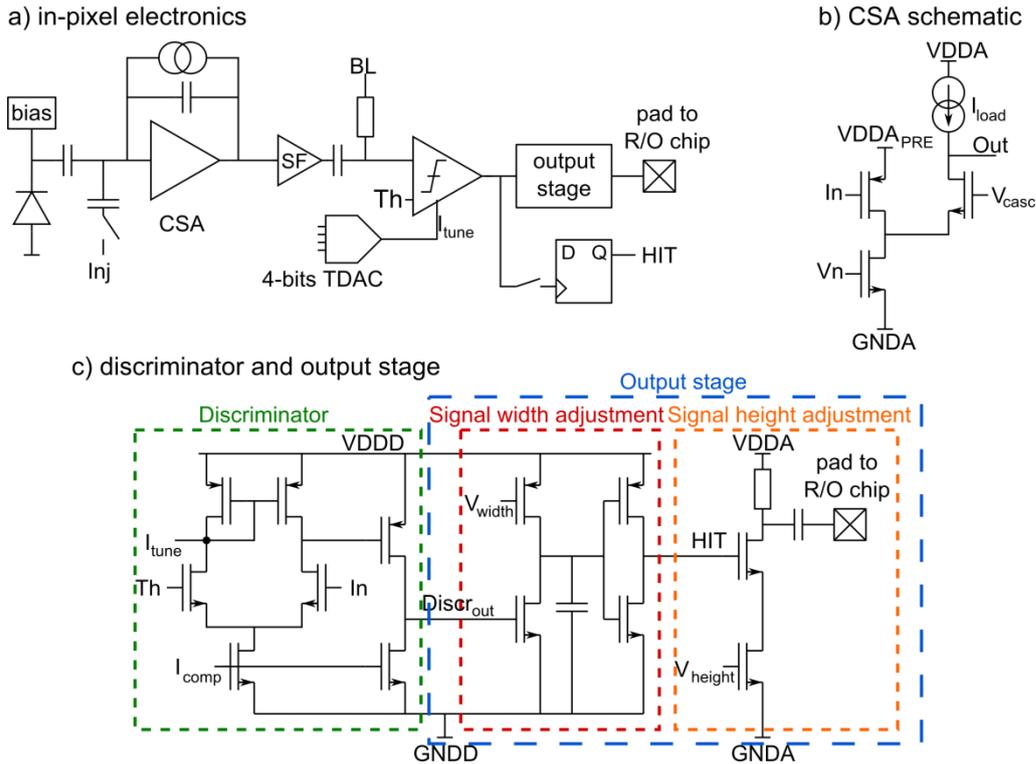

**Figure 2.** In-pixel electronics.

### 2.2.2 Chip composition

For versions A and B the chip architecture is primarily the same. The majority of the area is occupied by the pixel matrix, below which all global configuration DACs are placed. Pixels are configured (and read out in standalone mode) via a global shift register.

As mentioned in Section 2.1, the matrix comprises 2736 pixels, which are organised in 114 columns and 24 rows. Inside the matrix, pixels are organised into groups of three, and all pixels from the group are connected to one bond pad. This way a subpixel encoding of FEI-4 is achieved. For each CCPD_LF pixel in this three-pixel group, a different amplitude of output signal is assigned, so it can be distinguished which pixel fired.

Although the sensor structure for all pixels in the matrix is the same, for the readout electronics a few flavours have been implemented. All flavours are based on the circuit described in Section 2.2.1, with only small differences between them. In chip version A three kinds of pixels are present differing by the length and design (standard linear geometry or ELT – enclosed layout geometry) of transistor in CSA's feedback loop. In version B there are twelve pixel flavours, the differences between them include the length and design of transistor in CSA's feedback loop, method of high voltage delivery (through diode or resistor), and amount of deep P-well in the pixels. Distribution of pixel flavours in the matrices is shown in Figure 3.

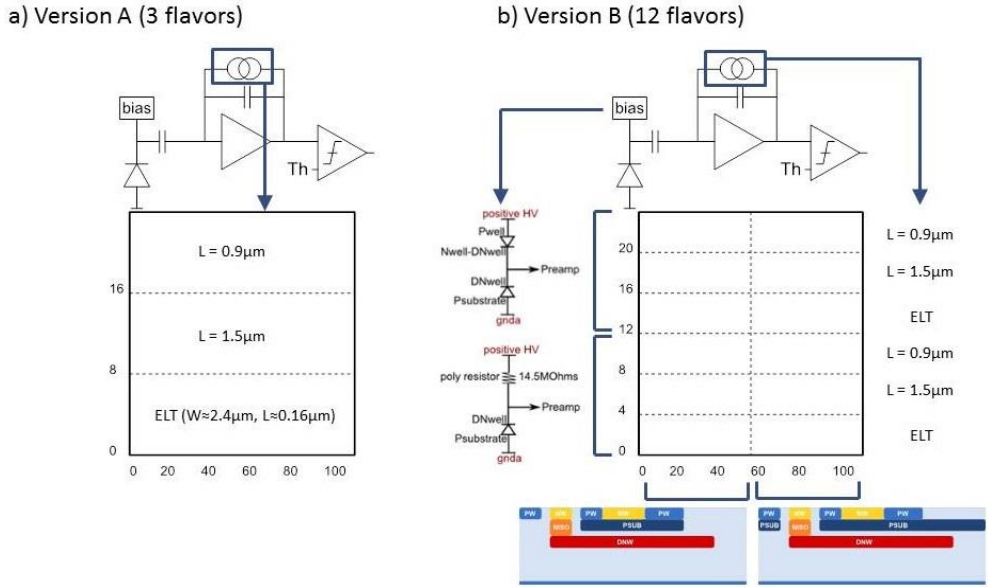

**Figure 3.** Pixel flavour distribution in version A and B matrices.

## 3. Measurement results

### 3.1 Breakdown voltage

As indicated by Equation 1, the depletion width is proportional to the square root of the reverse bias voltage. Therefore, it is beneficial to apply as high voltage as possible. This maximum value is defined by the breakdown voltage of the chip. The measured I-V curves for both versions are shown in Figure 4. For version A the breakdown voltage was estimated as 115 V. In version B the way the circuit is built limits the maximum applicable depletion voltage, because there is a capacitor between the collection well and the CSA input, which is connected to high voltage. As a result, before the real junction breakdown can occur, the dielectric of the mentioned capacitor will break and damage the chip. For this reason, the measurement was done only up to 26 V, which is below the breakdown voltage of the chip. This shows the shortcoming of the architecture implemented in version B.

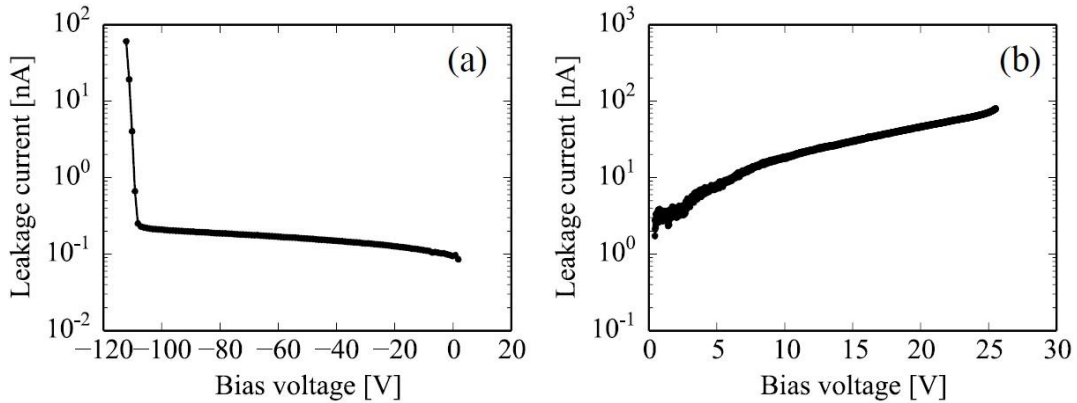

**Figure 4.** a) I-V curve of chip version A, b) I-V curve of chip version B.

## 3.2 Pixel performance

Pixel performance was first evaluated using an injection circuit, which, combined with the threshold scan method, allows measuring the gain and noise of the CSA. The result of such a procedure for chip version B (with injection equivalent to 3000e⁻) is shown in Figure 5. While gain values are rather uniform across the entire matrix and in agreement with circuit simulations, the noise map clearly indicates that the resistor-based high voltage bias introduces significantly more noise to the system than the diode-based one. This can be explained by the parasitic capacitance of the polysilicon bias resistor.

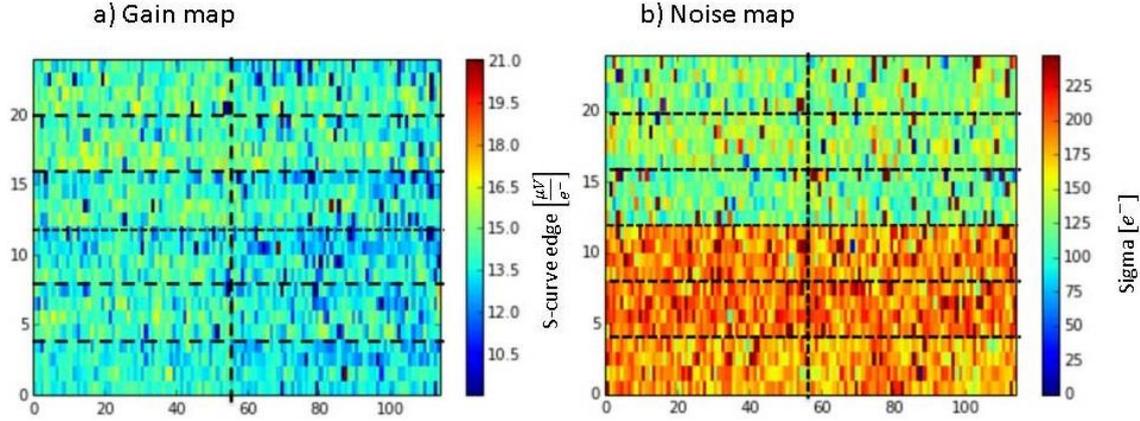

**Figure 5.** Gain and noise maps for version B. Black dashed lines indicate regions of different pixel flavours.

One of the main problems found regarding in-pixel electronics is a significant threshold dispersion of the discriminator. Figure 6a presents a histogram of the measured deviation of the discriminator's threshold from its average value, fitted with a Gaussian distribution. The sigma of this fit is $\sigma_{meas}$ = 11.7 mV, which means that the dispersion is too large to be corrected with a local 4-bit trim DAC ($LSB_{TDAC}$ = 0.78 mV). As a result, tuning the presented prototype to any given threshold value is problematic, which makes chip characterisation (e.g., charge collection efficiency) difficult. The described effect was attributed to the strong influence of the process variation on the discriminator's input transistors due to their small size and was reproduced in simulation (Figure 6b).

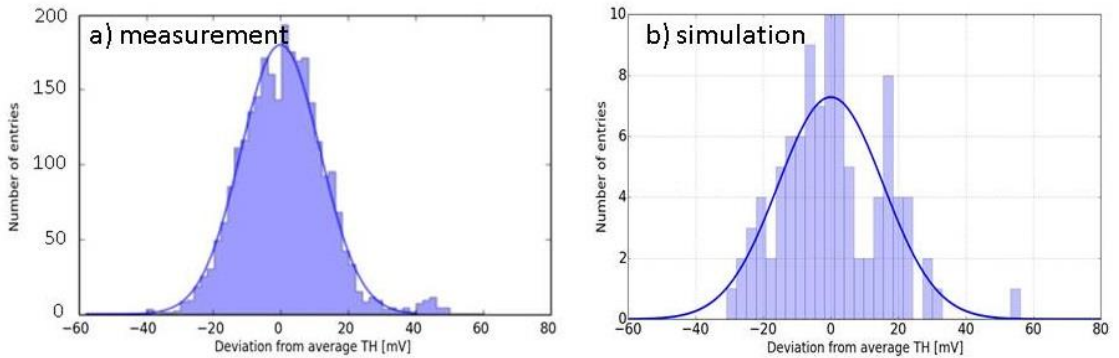

**Figure 6.** Histograms of deviation of the discriminator's threshold from it average value (blue bars) fitted with a Gaussian distribution (solid dark blue lines). Plane a) presents the measurement results from matrix version A, while histogram in Plane b) is a result of 100 Monte Carlo simulation runs of the pixel circuit (including extracted parasitic elements).

Two analogue buffers were integrated into the chips to allow monitoring of each pixel's CSA and discriminator output signals (one pixel at a time). This allowed the measurement of single pixel spectra. In Figure 7, the results of the front side illumination of chips with $^{55}$Fe are presented. The peak at 10.2 mV (version A) and 13.5 mV (version B) is a 5.9 keV $K_\alpha$ line from the source, the lower tail is an effect of charge sharing, and the peak at 2 mV is an artefact of the measurement setup. Both gain and noise values (6.2 $\mu V/e^-$, 149 e$^-$ ENC for version A and 7.9 $\mu V/e^-$, 120 e$^-$ ENC for version B) are comparable with simulation results. Differences in comparison with Figure 5 come from the fact that $^{55}$Fe spectra were obtained by observing signals through analogue buffers.

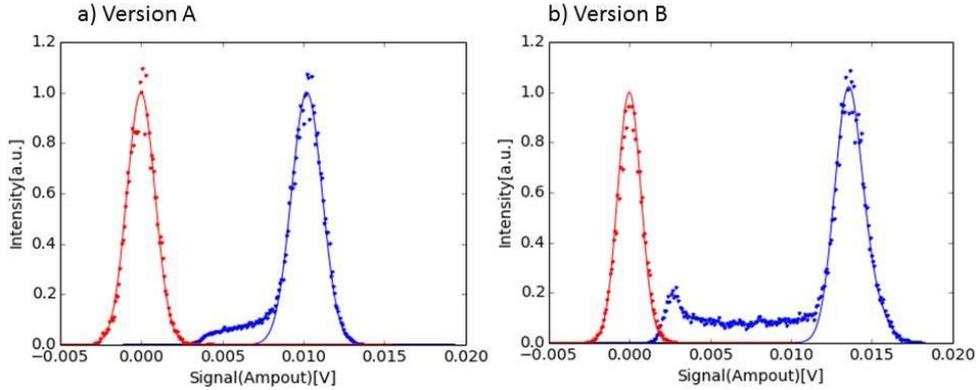

**Figure 7.** Single pixel spectrum for versions A and B. The baseline is shown in red, while signal from $^{55}$Fe is plotted in blue. Measurement points (dots) were fitted with the Gaussian distribution (solid lines).

### 3.3 Estimation of depletion width

The depletion depth was measured using a 3.2 GeV electron beam of an Electron Stretcher Accelerator (ELSA) at the University of Bonn [8]. First, a single pixel spectrum was calibrated using $^{55}$Fe, $^{109}$Cd, and $^{241}$Am sources. Afterwards, the spectra of the beam were measured for different values of reverse bias voltage. Obtained spectra were fitted with Landau-like functions to find the most probable value. Based on that, the maximal measured depletion depth was estimated [9] as 166 μm for version A (at 110 V) and 85 μm for version B (at 20 V). The measurement results are summarised in Figure 8, where it is shown that they match the analytical calculation for 3 kΩ·cm resistivity P-N diode.

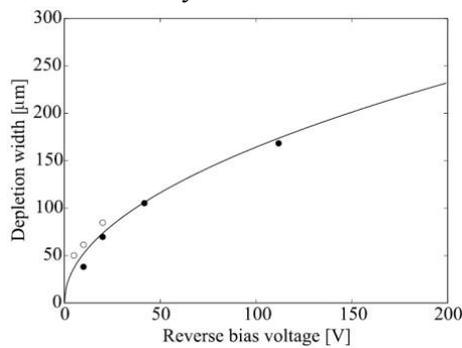

**Figure 8.** Estimated depletion depth as a function of reverse bias voltage. Measurement points are indicated by circles (filled – version A, empty – version B). Solid line is the calculated depletion width of substrate with 3 kΩ·cm resistivity.

### 3.4 Time walk

An electron beam was also used to measure the time walk. The prototype was put into the beam, and a scintillator was placed behind it. The outputs of CSA and output stage of a pixel were monitored via analogue buffers using an oscilloscope, while the scintillator signal was used as a trigger. For this measurement, time walk was defined as the dispersion of a delay time measured from the scintillator's signal to the edge of a monitored output stage response. The measurement for version B of the chip was carried out for two threshold settings, and the results are presented in Figure 9. For proper operation at ATLAS, the time walk must be smaller than 25 ns [10], any hits after that will be considered as noise. In the case of the discussed prototype the fraction of "in-time" hits was 79% for high threshold (2600$e^-$) and 91% for low threshold (190$e^-$). At the same time the noise occupancy measured for the low threshold was approx. 240 hits/mm/s, which significantly exceeds the maximum noise occupancy allowed in ATLAS ($10^{-6}$ hits/mm/s [10]). This shows that time walk is one of the most important issues to be fixed in the next chip iteration. Although the measurement was not repeated for chip version A, the expected result is worse than presented due to a larger ENC.

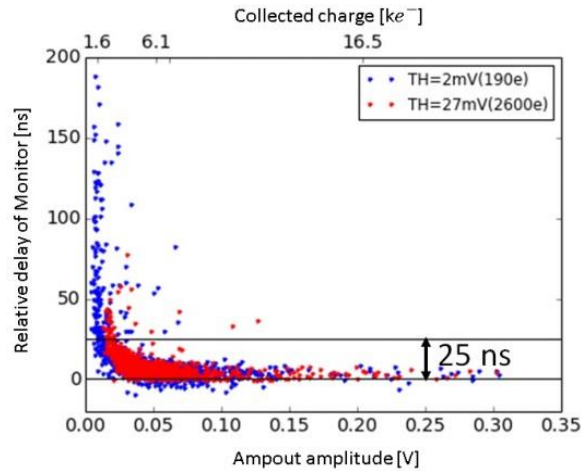

**Figure 9.** Time walk measurements for version B. The horizontal line marking 25 ns indicates the time limit for hits to be considered "in-time".

### 4. Plans for next prototypes

Based on the experience and knowledge gained from the measurements of the presented device, two more prototypes are being developed:

1. LF-CPIX is a direct successor of CCPD_LF using the same architecture as in version A (mainly due to depletion voltage limitations of version B described in section 3.1). The effort here is focused on fixing the discovered issues (large threshold dispersion of discriminator, too large time walk) and making a larger matrix that will be compliant with the CMOS Demonstrator requirements (1cm × 1cm). At the moment of writing this publication, the LF-CPIX is close to being complete and is expected to be submitted in the coming weeks.
2. LF-CPIX_M is a fully monolithic design using analogue electronics and a discriminator from LF-CPIX combined with column-drain architecture for pixel readout. A fully custom logic optimised for low-noise operation is used for the readout. At the moment of writing this publication, LF-CPIX_M is in the early development stage, but low-noise operation of the circuit seems to be achievable according to simulation. Expected submission for this circuit is in the beginning of 2016.

## 5. Summary

A prototype of DMAPS was designed in an LFoundry 150 nm CMOS process and manufactured on a high resistivity wafer. Extensive measurements were carried out using injection circuits (pixel electronics' gain and noise, matrix homogeneity), radioactive sources (charge spectra), and electron beams (depletion depth, time walk). Results are in agreement with simulation predictions. The device is the first step in confirming validity of the concept of a DMAPS particle detector in the context of the ATLAS Phase-II upgrade. Further studies, such as charge collection efficiency and influence of irradiation on performance, are currently ongoing. Issues found in this device will be addressed in the next iteration of prototypes, which is already under design.

## Acknowledgments

The authors would like to thank for the help provided by LFoundry (especially Gerhard Spitzlsperger and Gerhard Viertlmeister) and by the team of ELSA. This research project has been supported by a Marie Curie Initial Training Network fellowship of the European Community's Seventh Framework Programme under Contract number PITNGA-2011-289161 (TALENT).

## References

[1] I. Peric et al., *Overview of HVCMOS pixel sensors*, JINST **10** (2015) C05021

[2] B. Ristic, *Measurements on HV-CMOS active sensors after irradiation to HL-LHC fluences*, JINST **10** (2015) C04007

[3] M. Havránek et al., *DMAPS: a fully depleted monolithic active pixel sensor — analog performance*, 2015 JINST **10** P02013

[4] T. Oberman et al., Characterization of a Depleted Monolithic Active Pixel Sensor (DMAPS) prototype, JINST 10 (2015) C03049

[5] T. Hirono et al., *Characterization of CMOS Active Pixel Sensors on High Resistive Substrate*, Proceedings of 17[th] iWoRiD International Workshop on Radiation Imaging Detectors

[6] G. F. Knoll, *Radiation detection and measurements*, John Willey and Sons, Inc., New York (2000)

[7] http://www.lfoundry.com/

[8] W. Hillert, *The Bonn Electron Stretcher Accelerator ELSA: Past and future,* Europ. Phys. Jour. **A 28** (2006) 139

[9] S. Meroll et al., *Energy loss measurement for charged particle in very thin silicon layers*, JINST **6** (2011) P06013

[10] ATLAS Collaboration, *CERN Letter of Intent for the Phase-II Upgrade of the ATLAS Experiment*, Technical Report, Dec. 2012 CERN-LHCC-2012-022, LHCC-I-023